\documentclass[11pt]{article}

\usepackage[margin=1in]{geometry}

\usepackage{amsmath,amssymb,amsfonts,bm}
\usepackage{latexsym}
\usepackage{calc}
\usepackage{xspace}
\usepackage[usenames]{color}
\usepackage{xcolor}
\usepackage{graphicx}
\usepackage{setspace}
\newcounter{rot}%\addtocounter{rot}{1}, \therot

\definecolor{brown}{cmyk}{0, 0.72, 1, 0.45}
\definecolor{grey}{gray}{0.5}

 \def\b{\beta} \def\d{\delta} \def\D{\Delta}
\def\e{\epsilon}  \def\F{{\Phi}}  \def\g{\gamma}
\def\G{\Gamma}  
     \def\l{\lambda}
   \def\p{\pi}
\def\r{\rho}   
\def\t{\tau} \def\om{\omega}  \def\Om{\Omega}
\def\ve{\varepsilon}

\newtheorem{theorem}{Theorem}
\newtheorem{lemma}[theorem]{Lemma}
\newtheorem{corollary}[theorem]{Corollary}
\newtheorem{observation}[theorem]{Observation}

\newcommand{\proofstart}{{\bf Proof\hspace{2em}}}
\newcommand{\proofend}{\hspace*{\fill}\mbox{$\Box$}}

\def\cW{{\cal W}}

\newcommand{\ooi}{(1+o(1))}

\newcommand{\ul}[1]{\mbox{\boldmath$#1$}}

\newcommand{\rdup}[1]{{\lceil #1 \rceil }}
\newcommand{\rdown}[1]{{\left\lfloor #1\right \rfloor}}

\newcommand{\brac}[1]{\left(#1\right)}

\newcommand{\bfrac}[2]{\left(\frac{#1}{#2}\right)}

\newcommand{\rai}{\rightarrow \infty}

\def\sm{\setminus}
\def\seq{\subseteq}
\def\es{\emptyset}

\def\E{\mbox{{\bf E}}}

\def\Pr{\mbox{{\bf Pr}}}
\def\whp{{\bf whp}}

\newcommand{\ignore}[1]{}

\newcommand{\cA}{{\cal A}}

\newcommand{\beq}[1]{\begin{equation}\label{#1}}
\newcommand{\eeq}{\end{equation}}

\def\hpi{\widehat{\pi}}

\begin{document}

\title{
Random walks which prefer unvisited edges.
Exploring high girth even degree expanders in linear time
}

\author{
Petra Berenbrink\thanks{
School of Computing Science, Simon Fraser
University, Burnaby, Canada
} \and Colin Cooper\thanks{Department of
Informatics, King's College, London, U.K.} \and
Tom Friedetzky
\thanks{School of Engineering and Computing Sciences,
Durham University, Durham, U.K. } }

\date{\today}

\maketitle

\begin{abstract}
In this paper, we consider a modified random walk which uses
unvisited edges whenever possible, and makes a simple random walk otherwise.
We call such a walk an {\em edge-process} (or  $E$-process).
We assume there is a rule $\cal A$, which tells the walk which unvisited edge to use
whenever there are several unvisited edges.
In the simplest case, $\cal A$ is a uniform random choice over unvisited edges
incident with the current walk position.
However we do not exclude arbitrary choices of rule $\cal A$. For example, the rule could be determined
on-line by an adversary, or could vary from vertex to vertex.

For the
 class of  connected, even degree graphs $G$ of constant maximum degree, we characterize the vertex cover time of the $E$-process
in terms of the edge expansion rate of $G$, as measured by eigenvalue gap $1-\l_{\max}$ of the transition matrix of
a simple random walk on $G$.

A vertex $v$ is $\ell$-good, if any even degree subgraph containing all edges
incident with $v$ contains at least $\ell$ vertices.
A graph $G$ is  $\ell$-good, if every vertex has the $\ell$-good property.

In particular, for even degree expander graphs, of bounded maximum degree,
we have the following result.
Let $G$ be an $n$ vertex $\ell$-good expander graph.
Any $E$-process on $G$ has cover time
\[
C_G(E-\text{process})= O \left( n+ \frac{n \log n}{\ell}\right).
\]
This result is independent of the rule $\cal A$ used to select the order
of the unvisited edges,
which can be chosen on-line by an adversary.

With high probability random $r$-regular graphs, $r \ge 4$ even, are expanders
for which
$\ell = \Omega (\log n)$.
Thus, for almost all such graphs, the vertex cover time of  the $E$-process is $\Theta(n)$.
This improves the vertex cover time of such graphs by a factor of $\log n$,
compared to the $\Omega (n \log n)$ cover time of  {\em any } weighted random walk.
\end{abstract}

\newpage

\section{Introduction}

In a simple random walk on a graph, at each step a particle moves from its current vertex position to
a neighbouring vertex chosen uniformly at random.
Formally, a {\em simple} random walk $\cW_v=(\cW_v(t),t=0,1,\ldots)$ is defined as follows:
$\cW_v(0)=v$ and given $x=\cW_v(t)$, $y=\cW_v(t+1)$ is a randomly chosen neighbour of $x$.

\ignore{
If the walk makes a transition along the edge $\{x,y\}$, in either direction, we say
the edge $\{x,y\}$ is visited. If currently, the walk has  never made a transition along the edge $\{x,y\}$, we say
the edge  $\{x,y\}$ is unvisited. Initially all edges are unvisited, and the number of
unvisited edges can only decrease as the walk proceeds.
}

In this paper, we consider a modified  walk which uses unvisited edges whenever possible, and makes a simple random walk otherwise.
We call such a walk an {\em edge-process} (or  $E$-process).
At  each step the edge-process makes a transition to a neighbour  of the currently occupied vertex  as follows:
\begin{quote}
If there are {\em unvisited  edges} incident with the current
vertex pick one
and make a transition along this edge.

If there are no unvisited edges incident with the current vertex, move to a random neighbour
using a simple random walk.
\end{quote}
If we wish, can we assume there is a rule $\cA$, which tells the walk which unvisited edge to use
whenever there is a choice.
In the simplest case, this  is a uniform random choice over unvisited edges
incident with the current walk position.
However we do not exclude arbitrary choices of rule $\cA$. For example, the rule could be deterministic or decided
on-line by an adversary, or could vary from vertex to vertex.

The $E$-process seems particularly adapted to searching in a physical environment,
where edges can easily be marked as visited.
Imagine walking in a labyrinth, and  marking the entries and exits of the edges taken
with a piece of chalk. Whenever all exits are marked, walk randomly.

For any process which explores a graph $G$ by walking from vertex to vertex,
the {\em vertex cover time}, $C_G$, is defined as follows.
For $v\in V$, let $C_v$ be the expected time taken for a
walk $W$ on $G$ starting at $v$, to visit every vertex of $G$.  The
vertex cover time is defined as $C_G=\max_{v\in V}C_v$.
It was shown by
 Feige  \cite{Fe2},
that for any connected $n$-vertex graph $G$, the cover time of a simple random walk satisfies
$C_G \ge (1-o(1))n\log n$. In fact,  any weighted reversible random walk
has a lower bound on the cover time of $C_G= \Om( n \log n)$.
Thus no reversible random walk  can  have an $o(n \log n)$ cover time
A proof of the $\Om( n \log n)$ lower bound
on the cover time of weighted random walks, due to T. Radzik \cite{Rad},
 is given in Section \ref{weighted}.

%{\large\bf here it sounds that the bound is a contribution of the paper but the proof is by Tomasz}

One random process
similar to the $E$-process, is
the  {\em Random Walk with
Choice}, (RWC($d$)), of  Avin and Krishnamachari  \cite{A-K}.
The process
RWC($d$)
selects $d$ neighbours uniformly at random at each step, and  moves
to the least visited vertex among them. The paper \cite{A-K} makes an experimental study of
the process RWC($d$) on geometric random graphs, and the toroidal grid, and
finds reductions in cover time, and improved concentration of experimental results.
Recently a  special case of  the $E$-process has been studied by Orenshtein and Shinkar \cite{OS} in the context of edge cover times.
In \cite{OS}, the next unvisited edge is chosen u.a.r.
For a further discussion on edge cover time see below.

%{\large\bf I think here it is not clear where the rune A refers too. We did not define edge cover time so far, did we?}

In the context of deterministic walks,
the $E$-process has similarities with
the rotor-router, or Propp machine model; see
\cite{CDFS08} for an introduction to this topic.
The analysis of both processes depends on the underlying Eulerian properties of the
graph. In the case of the rotor-router process, the graph is turned into an Eulerian digraph
by replacing each edge with a pair of oppositely directed edges.
The vertex cover time of the rotor-router model is
$O(m D)$,
where $m$ is the number of edges of $G$, and $D$ is the diameter, see \cite{YWB03}.

The class of graphs we consider are connected, even degree graphs $G$ of constant maximum degree $\D(G)$.
We define a local expansion property
of vertices.  We say a vertex $v$ is $\ell$-good, if any even degree subgraph containing all edges
incident with $v$ contains at least $\ell$ vertices.
A graph $G$ is  $\ell$-good, if every vertex has the $\ell$-good property.
We  characterize the cover time of the $E$-process
in terms of the edge expansion rate of $G$, as measured by eigenvalue gap $1-\l_{\max}$ of the transition matrix of
a simple random walk on $G$.
A general statement of our result is the following theorem.

%{\large\bf It might be confusing that we sometimes use $\lambda_2 $ and sometimes $\lambda_{max}$.}
\begin{theorem}\label{Th1}
Let $G$ be a connected $n$ vertex even degree graph, with finite maximum degree,
and the additional property that 
that $G$ is $\ell$-good.
Then, {\bf any} $E$-process on $G$ has cover time
\[
C_G(E-\text{process})= O \brac{n+   \frac{n \log n}{\ell (1-\l_{\max})}}.
\]
\end{theorem}
%{\large\bf Here I had problems with the structure of the two lists. I think item 3 and
%6 of List 2 should be in List 1 since it refers to the Theorem. Why once numbers and once dots?
%Shal we maybe have 3 lists, one for the Propp model relation?}
We briefly list a series of remarks and corollaries which arise from Theorem \ref{Th1}
\begin{itemize}
\item[i)]
The upper bound on the cover time given in
 Theorem \ref{Th1}  is independent of the rule $\cA$ used to select unvisited edges,
even if  this choice is decided on-line by an adversary.

\item[ii)]
For expander graphs, which have positive constant
  eigenvalue gap, Theorem \ref{Th1} becomes
\begin{equation}\label{Th2}
C_G(E-\text{process})= O \brac{n+   \frac{n \log n}{\ell}}.
\end{equation}
In particular, 
for $\ell$-good  even degree expanders
where $\ell =\Om(\log n)$,
the $E$-process covers the graph in $\Theta(n)$ steps.
As any walk-based process must take $n$ steps to visit every vertex,
the order of our result is best possible.

\item[iii)]
Examples of $\ell$-good graphs  where
$\ell =\Om(\log n)$
 include  random $r$-regular graphs, for which we have the following corollary.
\begin{corollary}\label{rreg}
Let $r \ge 4$
even. Let ${\cal G}_r$ denote the class of random $r$-regular graphs.
Let $G$ be sampled uniformly at random from ${\cal G}_r$, then with high probability
$C_G(E-\text{process })=O(n)$.
\end{corollary}
See Section \ref{EXC} for the proof of this.
Other examples of $\ell$-good graphs are
 random graphs with fixed degree sequence $\ul d$, and all vertices of even degree at least 4,
(ii) algebraically constructed even degree expanders of
logarithmic girth, see \cite{LPS}.

\item[iv)]
The lower bound on the cover time of $G$
by {\em any} weighted reversible random walk is  $\Om(n \log n)$.
(See Section \ref{weighted} for a proof of this  result).
For expanders, the comparable cover time is given by \eqref{Th2}.
Up to $\ell = \log n$, this gives a speed up
of $\Om(\log n/\ell)$ compared to {\em any} random walk.

\item[v)]
In Section \ref{odd-deg} we give some experimental results on the performance of the $E$-process.
Simulations suggest that for even degree random regular graphs, the cover time
of the $E$-process
is bounded (asymptotically) by the number of edges $m$ in the graph (see Figure \ref{fig:covertime}).

Could we expect an $O(n)$ cover time for the $E$-process on odd  degree expanders?
Experimentally, we find that this is not the case (see Figure \ref{fig:covertime}).

\item[vi)]
A practical consequence of Theorem \ref{Th1}, is that, in order to build \lq easy to search\rq\ networks,
 we should ensure all vertices have even degree and few short cycles. Examples of such constructions,
 based on  even degree random $r$-regular graphs, are the SWAN  P2P network of \cite{swan} based on switches, and the
 flip based P2P network of \cite{MS}.  Properties of these networks
 such as  connectivity, diameter and mixing-rate
  were studied in  (e.g.)
 \cite{CDG},\cite{CDH},  \cite{FGMS}.

\end{itemize}
We also make some observations on edge cover time of the $E$-process (see i,ii below), and on the
relationship between the $E$-process and Propp machines (items iii-v).
\begin{itemize}
\item[ i)]
In general
 upper bounds on the edge cover time of the $E$-process
 depend on the number of short cycles.
 The girth $g$ of a graph  $G$ is  the minimum  length  cycle in $G$.
It can be shown that the
 $E$-process will cover all edges of a connected even degree graph in $O(|E|+ n \log n /(1-\l_2)^2 g)$.
This bound can be improved if the number of short cycles can be upper bounded.
As an example, for even degree random regular graphs, the (\whp) upper bound on  the edge cover time
is $O(n \om)$, where $\om \rai$ arbitrarily slowly.
%{\large\bf I can not understand the last sentence}

\item[ii)]
The result of \cite{OS}
gives a bound for
edge cover time of $r$-regular
graphs  of $O(|E|+ n \log n /(1-\l_{\max}))$.
This is
at best  $O(n \log n)$ for sparse graphs, but is tight for expanders
 provided the number of edges $|E| =  \Om(n\log n)$.
This result differs qualitatively from Theorem \ref{Th1} which
 treats vertex cover time of  constant degree expanders ($|E|=cn$, $c$ constant).

\item[iii)]
Suppose it is the case that the edges of a graph $G$ can be distinguished
as unvisited in each direction by the $E$-process; i.e. a
first visit $(x,y)$ and a first visit $(y,x)$ are regarded as distinct.
This converts $G$ into an Eulerian digraph, so that
 the even degree restriction  is no
longer necessary, and Theorem \ref{Th1} now holds for all connected graphs
of bounded degree.

\item[iv)]
Suppose the  edges of the graph can be  marked as unvisited
in each direction. Then the ordering of the (directed) unvisited edges at each vertex made by the rule $\cA$
is a rotor order for a rotor-router (Propp machine).
The $E$-process acts as  a hybrid of
a Propp machine and a random walk, the algorithm being: {\em Use the rotor once at each vertex
and then walk randomly. Any rotor order will do}. The power of the adversary is to set
the rotor order.

\item[v)]
In some rotor-router models an adversary can force a cover time of
$\Om(m \log m)$ on  connected $m$ edge graphs (see \cite{BGN} for details).
This phenomena partially arises  because the adversary can make the walk  retrace
visited edges, even when unvisited edges are present at a vertex. In the $E$-process  the adversary is less strong, and
only has power to select the next unvisited edge used by the process. All transitions over visited edges are chosen randomly.
Thus when $m=\Theta(n)$, and $\ell=\Omega(\log n)$,  the $E$-process has cover time $\Theta(n)$,
as compared to  $\Theta(n \log n)$ cover time in the aforementioned adversarial rotor-router model.
%Also, the random walk component of the $E$-process  destroys the power of the adversary
%to determine the order in which to retrace visited edges.

\end{itemize}

\subsection{Random walk properties}\label{RWP}

Let $G=(V,E)$ denote a connected graph, $|V|=n$, $|E|=m$, and let
$d(v)$ be the degree of a vertex $v$.
% and let $N(v)$ be the  neighbourhood of $v$.
A {\em simple random walk} $\cW_u,\,u\in V$, on graph $G$ is a
Markov chain modeled by a particle moving from vertex to vertex
according to the following rule. The probability of transition from
vertex $v$ to  vertex $w$ is equal to $1/d(v)$, if $w$ is a
neighbour of $v$, and $0$ otherwise. The walk $\cW_u$ starts from
vertex $u$ at $t=0$. Denote by $\cW(t)$ the vertex reached at step
$t$; $\cW(0)=u$.

Let $P$ be the transition matrix of a simple random walk on a graph
$G$. Thus  $P_{i,j}=1/d(i)$ if and only if there is an edge between $i$ and $j$
in $G$. Let $P_{u}^{(t)}(v)=\Pr(\cW_{u}(t)=v)$ be the $t$-step
transition probability. We assume the random walk $\cW_{u}$ on $G$
is ergodic with stationary distribution $\pi$, where
$\pi_v=d(v)/(2m)$. If this is not the case, e.g. $G$ is  bipartite,
then the walk can be made ergodic, by making it lazy. A  random walk
is {\em lazy}, if it moves from $v$ to one of its neighbours $w$
with probability $1/(2d(v))$, and stays where it is (at vertex $v$)
with probability $1/2$.

Let $1, \l_2,...,\l_n$, be the  eigenvalues of $P$, and let
$\l_{\max}=\min(|\l_2|,|\l_n|)$.
We henceforth assume that $\l_2=\l_{\max}$ which can be achieved by
making the chain lazy. This has no significant effect on our
analysis.

The convergence to stationarity of a simple random walk is bounded by
\begin{align}\label{mix}
&|P_{u}^{(t)}(x)-\pi_x| \leq (\p_x/\p_u)^{1/2}\l_{\max}^t .
\end{align}

\paragraph{Visits to a Single Vertex}

For a random walk starting from vertex $u$, let $H_v$ be the number of steps
taken to reach vertex $v$, and let $\E_u(H_v)$ be the expected value of $H_v$;
the expected hitting time of $v$ starting from $u$. If the distribution of
the random walk at some step is $\r=(\r(u), u \in V)$, we can similarly define
the hitting time from starting distribution $\r$ as
$\E_\r(H_v)=\sum_{u \in V} \r(u) E_u(H_v)$.

For a random walk starting at a vertex chosen from the stationary distribution $\pi$,
let $\E_{\pi}(H_v)$ denote the expected hitting time of vertex $v$ from stationarity. The quantity $\E_{\pi}(H_v)$
 can be expressed in the following way, (see e.g.~\cite{AlFi}, Chapter 2)
\begin{equation}\label{pi-hit}
\E_{\pi}(H_v) = Z_{vv}/\pi_v,
\end{equation}
where
\begin{equation} %\label{pihitz}
Z_{vv} = \sum_{t=0}^{\infty} (P_{v}^{(t)}(v) - \pi_v).
\end{equation}
Using \eqref{mix}, we can bound the value of $\E_{\pi} (H_v)$
as follows.

\begin{lemma}\label{Eval-gap}
\begin{equation}\label{hitit-eval}
\E_{\pi} (H_v) \le \frac{1}{(1-\l_{\max}) \pi_v}.
\end{equation}
\end{lemma}
\proofstart Using \eqref{mix} with $x=u=v$, then
\[
|P_v^t(v)-\pi_v| \le (\l_{\max})^t,
\]
and
\[
Z_{vv}=\sum_{t \ge 0} (P_v^t(v)-\pi_v) \le  \sum_{t \ge 0} (\l_{\max})^t
= \frac{1}{1-\l_{\max}}.
\]
\proofend

Let $T_G$ be the mixing time of a graph $G$,
 such that, for $t\geq T_G$,
\begin{equation}\label{4}
\max_{u,x\in V}|P_{u}^{(t)}(x)-\pi_x| =O\bfrac{1}{n^3}.
\end{equation}
Let $\cA_t(v)=\cA_{t,u}(v)$ denote the event that $\cW_u$ does not
visit vertex $v$ in steps $0,...,t$. Lemma~\ref{crude} gives a
 bound for $\Pr(\cA_t(v))$ in terms of
$\E_{\pi}(H_v)$ and the mixing time $T$.

\begin{lemma}\label{crude}
Let $T_G$ be the mixing time of a random walk $\cW_u$ on $G$ satisfying
\eqref{4}. Then
\[
\Pr(\cA_t(v)) \le  e^{-\rdown{{t}/{(T_G+3\E_{\pi}(H_v))}}}.
\]
\end{lemma}

\proofstart Let $\r =(\r_w)$ be the distribution of $\cW_u$
on $G$ after $T=T_G$ steps,  where $\r_w= P_u^{(T)}(w)$.
Let $\E_{\r}(H_v)$ be the expected time
to hit $v$ starting from $\r$.
As $T$ satisfies \eqref{4}, and  $\pi_x=\Om(1/n^2)$ for any
connected graph, then $\r_w=\ooi \pi_w$.
It follows  that
\beq{this-eqn}
\E_{\r}(H_v)=\ooi \E_{\pi}(H_v).
\eeq
Let $H_v(\r)$ be the time to hit $v$ starting from $\r$, then
\[ \Pr[H_v(\r) \ge 3 \E_{\pi}(H_v)] \le \frac{1
}{e}.
\]
Let $\t=T+3\E_{\pi}(H_v)$.
By considering the process  $\cW_u$ at $\cW(0)=u, \cW(\t), \cW(2\t),
\ldots, \cW(\rdown{t/\t}\t)$ we obtain
\[
\Pr(\cA_t(v)) \le  e^{-\rdown{t/\t}}.
\]
\proofend

\paragraph{Visits to Vertex Sets}

We can extend the results presented above to any nonempty subset $S$ of vertices
in the following way. From $G$ we obtain a (multi)-graph
$\G=\G_S$ by contracting $S$ to a single vertex $\g$. Note that we
retain  multiple edges and loops in $\G_S$, so that $d(S)=d(\g)$,
and $|E(\G)| = |E(G)| = m$. Let ${\hpi}$ be the stationary
distribution of a random walk on $\G$. If $v \not \in S$ then
$\hpi_v = \pi_v$, and $\ \hpi_{\g}=\pi_S \equiv \sum_{x\in S}
\pi_x$.

For $u \not \in S$
let $\cW_u$ be a walk starting from $u$ in $G$, and let
 $\widehat{\cW}_u$ be the equivalent walk starting in $\G$.
Provided $\cW_u$ does not visit $S$ in $t$ steps, (the event
$\ul A_t(S,G)$),  then $\widehat{\cW}_u$
does not visit $\g$ (the event $A_t(\g,\G)$), and the walks have the same transition
probabilities. Thus,
\[
\Pr(\ul A_t(S,G))=  \Pr(\ul A_t(\g,\G)),
\]
and
\begin{align}\label{eq:expectedHittingSet}
\E_{\pi}(H_S) \; = \; \E_{\hpi}(H_\g).
\end{align}

It is a known result that contracting vertex sets increases the eigenvalue gap.
(For a proof see e.g. \cite{AlFi} Chapter 3, Corollary 27.)
Thus
\[
1-\l_{\max}(G) \le  1-\l_{\max}(\G).
\]
In our proofs, we will always choose a mixing
time $T$ in \eqref{4} satisfying both $T \ge T_G$, and $T \ge T_{\G}$.
It follows that, using this mixing time $T$, the results of Lemma \ref{Eval-gap}, and Lemma \ref{crude}
apply equally to $\G$, and to $G$. Thus e.g.
\begin{corollary}\label{HGam}
Let $G=(V,E)$, let $|E|=m$. Let $S \seq V$, and let $d(S)$ be the degree of $S$. Then $\E_{\pi} H_{S}$,
the expected hitting time of $S$ from stationarity satisfies
\[
\E_{\pi} H_{S} \le \frac{2m}{ d(S)(1-\l_{\max}(G))}.
\]
    \end{corollary}

\section{Proof of main result}

\subsection{Properties of the edge-process}\label{proces}

It is helpful to think of the progress of the $E$-process as a re-colouring
of the edges of the
graph $G$. We consider unvisited edges as coloured blue, and explored edges
as coloured red.
Let $X(t)$ be the position at step $t$ of a particle moving according to an $E$-process.

Initially, the
particle is at $X(0)=u$, the start vertex,
and all edges of the graph $G$ are coloured blue (unvisited).
Given $X(t)=v$, $X(t+1)$ is chosen
as follows.  If all edges incident with $v$ are red
(previously visited) the walk chooses $X(t+1)$ u.a.r. from $N(v)$.
If however, there are any blue (unvisited) edges incident with $v$, then we pick a
blue edge $(v,w)$ according to the rule $\cA$.  The walk then moves to $X(t+1)=w$, and
re-colours the edge $(v,w)$ red (visited).  We assume that the edge
$(v,w)$ is re-coloured red at the start of step $t+1$,  the instant at which
the walk arrives at $w$. Thus we regard the transition $(v,w)$ as being along a blue edge.

At each $t$ the next transition is either along a blue or a red edge. We speak
of the sequence of these edge transitions as the blue (sub)-walk and the red
(sub)-walk. The
walk thus defines red and blue phases which are maximal sequences of edge transitions when the
walk is the given colour.
For any vertex $v$, and step $t$, the blue (resp.~red) degree of
$v$ is the number of blue (resp.~red) edges incident with $v$ at the start of  step $t$.

\begin{observation}\label{obs1} Assume all vertices of $G$ are of even degree.
  Then a blue phase of the $E$-process which starts at a vertex $v$ (at some
  step $t$), must end at $v$ (at some step $t+\tau$).
\end{observation}

\proofstart
This follows from a simple parity argument.  The first blue phase starts at
$t=0$,  at the start vertex $u$.  At $t=0$  every vertex has even blue degree.
Suppose that at step $t$ we have $X(t)=w$, where $w \ne u$.  Inductively every
vertex, apart from the start vertex $u$ and the current position $w$ have even
blue degree, whereas the blue degree of $u$ and $w$ is odd, and hence greater than zero.
The particle can thus exit $w$ along a blue edge.
When the particle
leaves $w=X(t)$ making the transition $(X(t), X(t+1))$, then the blue degree
of $w=X(t)$ becomes even. If $X(t+1)=u$, then the degree of $u$ is even
and the particle has returned to the start. If $X(t+1) \ne u$, then the blue
degree of $X(t+1)$ and $u$ is odd.

If the particle returns to $u$ at step $t$, and the blue degree of $u$ is
zero, then the blue phase  at $u$ is completed at (the start of) step $t$.
The particle  now leaves $u$ along a red edge
$(u,v)=(X(t),X(t+1))$, and this is the beginning of a red phase.  Inductively,
the blue degree of $v$ is even when the particle arrives at $v$.  If
$v$ has blue edges incident with it, then a blue phase begins. Otherwise the
red phase continues.
\proofend

Note that it is
possible for all edges incident with  a vertex $v$
to be coloured red by transitions made during the blue sub-walk,
and that $v$ has not been visited by a red walk.

Let $G[S]$ denote the subgraph of $G$ induced by the set of vertices $S \seq V$.
The following summarizes the consequences of Observation \ref{obs1}.
\begin{observation}\label{subgraph}
Assume vertex  $v$ is unvisited at step $t$,
and that the $E$-process is in a red phase.
\begin{enumerate}
\item All edges incident with $v$ are blue at step $t$.
\item The blue degree of all vertices at step $t$ is even.
\item Let $S^*_v$ be the maximal blue  (unvisited), edge induced  subgraph obtained by fanning out
in a breadth first manner
from $v$ using only blue edges. Let $U^*$ be the vertex set of $S^*_v$. Then
\begin{enumerate}
\item  The degree of $v$ in $S^*_v$ is $d(v)$, the degree of vertex $v$ in $G$.
All vertices of $S^*_v$ have positive even degree.
\item All edges between $S^*_v$ and $G \sm U^*$ are red.
\item $G[U^*]$ may induce red edges, but these are not part of $S^*_v$.
\end{enumerate}
\end{enumerate}
\end{observation}
%{\large\bf Do we have to define $G[U^*]$?}
In the simplest case $S^*_v$  consists of $d(v)/2$ blue cycles with common  root vertex $v$,
but otherwise vertex disjoint.

It follows from Observation \ref{obs1}, that if we ignore
the blue phases of the $E$-process, then the resulting red phases describe a
continuous simple random walk $W_u(t_R)$ on the graph $G$. Each step $t_R$ of
the walk $W_u$ corresponds to some step $s > t_R$ in the $E$-process.  From
Observation \ref{obs1} it also follows that, if $X$ starts at $u$, then $W_u$ also
starts at vertex $u$.

At step $t$ of the $E$-process, we have $t=t_R+t_B$, where $t_R, \; t_B$ are the (unknown) number of red and blue edge transitions.
One thing is certain however; the length of the blue walk can be at most
the number of edges $m$ of $G$. This is formalized in the next observation.
\begin{observation}\label{gotall}
Let $W_u(t_R)$ be a simple random walk on the graph $G$ defined by the red phase of the $E$-process,
and let $X_u(t)$ be the  walk defined by the $E$-process. Then
$t_R  < t < t_R+ m$.
\end{observation}

%%%%%%%%%%%%%%%%%%%%%%%%%%%%%%%%%%%%%%%%%%%%%%%%%%%%%%%%%%%%%%%%%%%%%%%%%%%%%%%%%%%%%%%%%%%%%%%%%%%%%%

\subsection{Cover time of the $E$-process}\label{proof}\label{nice}

\begin{lemma}\label{bits}
Let $\cW_u$ be a random walk starting from $u$ in $G$.
 Let $S$ be a set of vertices of $G$ of
size $s$. Let $$d(S)=o(m/\log n),$$ where $d(S)$ be the sum of the degrees of the
vertices in $S$.
Let $$t = \Om (m/s(1-\l_{\max}),$$ then
\[
\Pr(S \text{ is unvisited by } \cW_u \text{ at step } t) =O\brac{e^{- t d(S)  (1-\l_{\max})/14m}}.
\]
\end{lemma}

\proofstart
Contract $S$ to a single
vertex $\g=\g(S)$, retaining all resulting loops and parallel edges. Denote
the resulting graph by $\G$. Let $|S|=s$.

For $\l \le 1$, $\l \le e^{-(1-\l)}$.
 It follows from \eqref{mix},  for given $u,x$ that
\begin{equation}\label{mixT}
|P_u^t(x)-\pi_x| \le \D^{1/2}e^{-(1-\l_{\max})t},
\end{equation}
where $\D$ is the maximum degree in $G$ or $\G$ as appropriate.
In either case, $\D\le 2m =O(n^2)$.
Let $$T=K \log n /(1-\l_{\max}),$$ where $K \ge 6$. As  there are at most $n^2$ pairs $u,x$,
then using \eqref{mixT}
\[
\sum_{u,x}|P_u^t(x)-\pi_x| \le n^2\D^{1/2} e^{- T(1-\l_{\max})} = O(1/n^3).
\]
Thus $T$ is a mixing time satisfying \eqref{4} in both $G$ and $\G$.
Also, from Corollary \ref{HGam} we have
\[
\E_{\pi}(H_S) \le \frac{2m}{d(S) (1-\l_{\max})}.
\]
For $u \not \in S$
let $\cW_u$ be a walk starting from $u$ in $G$, and let
 $\widehat{\cW}_u$ be the equivalent walk starting in $\G$.
Provided $\cW_u$ does not visit $S$ in $t$ steps, (the event
$\ul A_t(S,G)$),  then $\widehat{\cW}_u$
does not visit $\g$ (the event $A_t(\g,\G)$), and the walks have the same
probabilities. Thus
\[
\Pr(\ul A_t(S,G))=  \Pr(\ul A_t(\g,\G)).
\]
From Lemma \ref{crude} we have
\[
\Pr(\ul A_t(\g)) \le \exp\brac{-\rdown{{t}/{(T+3\E_{ \hpi}(H_\g))}}}.
\]
Let $T_\G$ be a mixing time of the random walk on $\G$ satisfying \eqref{4}.
From \eqref{mixT}, and
the conditions on $t, d(S)$ given in the lemma, we have  that $T_\G=o(m/d(S)(1-\l))$,
%{\large\bf did we define $T_\G$?}
and thus
\[
T+3\E_{ \hpi}(H_\g) \le \frac{7m}{d(S)(1-\l_{\max})}.
\]
We have the  result that
\[
\Pr(\ul A_t(S,G))
\le \exp\brac{-t \frac {d(S) (1-\l_{\max}) }{14m}}.
\]
\proofend

\begin{lemma}\label{numberof}
Let $G$ be a graph of maximum degree $\D$.
Let $\b(s,v)$ be the number of  connected edge induced subgraphs
of size $s$ rooted at  vertex $v$ in $G$. Then $$\b(s,v) \le 2^{s\D}.$$
\end{lemma}

\proofstart
We make a crude estimate for $\b(s,v)$ by building a digraph $H_v$
in a breadth first manner as follows.  Initially  $H_v=\es$ and all adjacent
edges of $v$ are  in $G$ are labeled unvisited.  Mark $v$ as processed and
add it to $H_v$.  For each edge incident with $v$, we label it as retained or
excluded.  Starting from $v$ there are $d(v)$ unvisited edges, and so at most
$2^{d(v)}$ choices for the subset of edges incident with $v$ to retain.
We process  each retained  edge $(v,u)$ in increasing endpoint label order.  Mark
$u$ as processed and add the retained edge  $(v,u)$ to $H_v$.  There are at
most $2^{d(u)-1}$ choices for labels (retained, excluded) of any unvisited edges
incident with $u$.

Thus we fan out from $v$ in a breadth-first manner using only retained edges,
$(u,w)$.  We add $w$ to $H_v$, and also any retained edges $(x,w)$, where $x$
was processed earlier than $w$.  In general there are some number of retained
and excluded edges incident with $w$ in $G$, resulting from processing earlier
vertices; and the remaining at most $(d(w)-1)$ edges are unvisited.
We continue until $H_v$ has  $s$ processed vertices, and the choices at these
vertices have been evaluated.  The $s$ processed vertices of $H_v$ and any
retained edges between them defines a connected subgraph of size $s$ rooted at
$v$, and every subgraph of size $s$ rooted at $v$ is found by this
construction.
\proofend

\begin{lemma}\label{proofof}
Let $G$ be an $\ell$-good  graph of minimum degree $\d$ and  maximum degree $\D$.
With probability $1-O(n^{-3})$, after
\[
\t^*= O\brac{ m \brac{1 + \frac{\D \log n}{\d\min(\ell, \log n)(1-\l_{\max})}}}
\]
 steps of the
$E$-process, no vertex of $G$ remains unvisited. The value of $\t^*$ is
independent of the choice of rule $\cA$ used by the process.

In particular, if $G$ has constant maximum degree, there exists a constant $B>0$ such that
 \[
 \t^*= B n [1+  (\log n)/\min(\ell, \log n)(1-\l_{\max})].
 \]
\end{lemma}

\proofstart
Let $S_v^*$ be the  maximal connected even degree blue
subgraph rooted at $v$, as described in Observation \ref{subgraph}.
Let $S_v$ be any connected subgraph of $S_v^*$ of size $$s=\min(\ell, \log n),$$ rooted at $v$.  By
Lemma \ref{numberof}, there are at most $2^{\D s}$ such possible subgraphs.

For a random walk $\cW_u$ starting from vertex $u$,
let $P(s,t)$ be the probability  that at step $t$ there exists an unvisited connected subgraph of size $s$
rooted at some vertex $v$. Thus using Lemmas \ref{bits} and \ref{numberof}
\[
P(s,t)\le n 2^{\D s} e^{- t \frac{d(S) (1-\l_{\max}) }{14 m} }.
\]
As $s =\min (\ell, \log n)$, on choosing
\[
t^*= (\D+7)   \log n \;\frac{14 m}{\d s(1-\l_{\max})},
\]
where $\d\ge 2$ is minimum degree, we find that
\begin{equation}\label{sick}
P(s,t^*)= O(1/n^3).
\end{equation}
From Observation \ref{gotall}, the length of the $E$-process walk on unvisited
edges is at most $m$, the number of edges of $G$, and
the step $\t^*=\t(t^*)$ in the $E$-process corresponding to the step $t^*$
in the red phase random walk $\cW_u$
is bounded by $\t^* \le m +t^*$. In particular,  if $\D$ is constant then $m= c n$, and
\[
\t^* \le m +t^*= B(n +(n \log n)/(\min(\ell, \log n) (1-\l_{\max}))).
\]
Suppose some vertex $v$ is unvisited at $\t^*$. Then
a blue (unvisited) edge induced subgraph  $S_v^*$ rooted at $v$ exists at $\t^*$.
However, from \eqref{sick},  \whp\ any $S_v \seq S_v^*$
of size $s$, contains a vertex $z$ already visited by $\cW(t^*)$.
Suppose this visit occurs at $t \le t^*$, but that, at step $t^*$,
 some edges incident with $z$ are unvisited, a necessary condition for $z \in S_v^*$.
On arriving at $z$, the $E$-process
completes the exploration of all edges incident with $z$,
after which the random walk $\cW(t)$ continues up to step $t^*$.
Thus at $\t^*$ all edges adjacent to $z$ are red, which is a contradiction.
\proofend

\begin{figure*}%[!ht]
  \centering
 \includegraphics{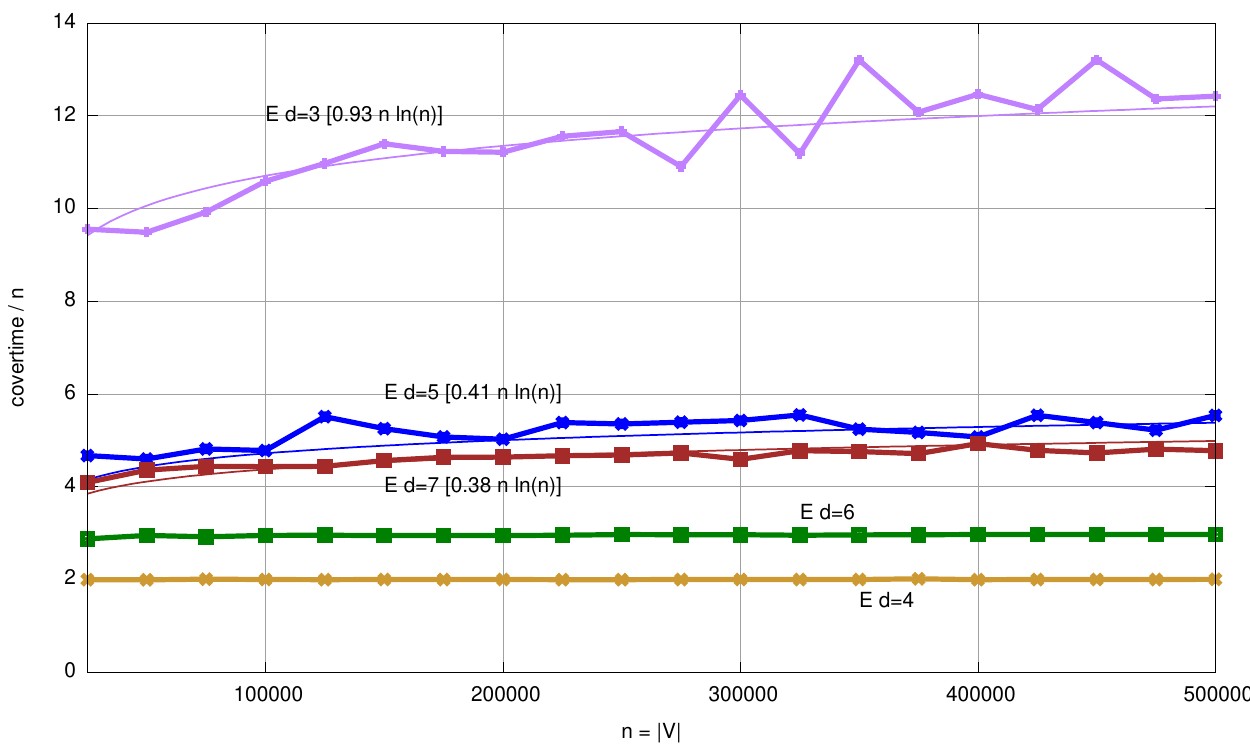}
   \caption{Normalised cover time of $E$-process as function of size and degree $d$}
  \label{fig:covertime}
\end{figure*}

%\section{Examples and counter in\-di\-ca\-tions}\label{EXC}
\section{Discussion and examples}\label{EXC}

%{\huge\bf Too long. no idea what to do}

\subsection{Lower bound cover time for weighted random  walks}\label{weighted}
For an introduction to properties of weighted random walks see \cite{AlFi}.
The following proof  that the cover time of any weighted random walk is $\Om(n \log n)$,
is due to T. Radzik \cite{Rad}.

For any vertex $u$, the expected first return time  $\E T^+_u$ to
$u$ is $\E T^+_u=1/\pi(u)$.

The commute time  $K(u,v)$ between vertices $u$ and $v$, is the
expected time  taken to go from vertex $u$ to vertex $v$ and then
back to vertex $u$. Formally,  $K(u,v)=\E_uT_v + \E_vT_u$.
Any walk starting from $u$ either visits $v$ on the way back to $u$ or it does not.
Thus  $\E T^+_u$ is at most the commute time  $K(u,v)$ between $u$ and $v$.

Let $S$ be the subset of vertices with $\pi(u) \le 2/n$. Thus $|S| \ge n/2$.
This follows because $\sum_{u \in V} \pi(u)=1$.
As $\E T^+_u=1/\pi(u)$, it follows that for $u \in S$,  $\E T^+_u \ge n/2$.

Let $K_S=min_{i,j \in S} K(i,j)$ then, $K_S\ge \E T^+_u \ge n/2$.
From \cite{KKLV}, we have the lower bound that
\[
 C_G \ge  (\max_{S \seq V} K_S \log |S|)/2 \ge (n/4) \log (n/2).
\]

\ignore{
{\bf\large I think here the problem I had was that we did not define commute time and I assumed it is the time to go from $u$ to $v$ only
I am fine with the proof, we might add de definition of commute time and return time}
Added defn
}

\subsection{Proof of Corollary \ref{rreg}}

Random $r$-regular graphs, ${\cal G}_r$,  with $r \ge 4$
even, are an example of a class of graphs for which (\whp)
$C_G(E-\text{process })=O(n)$.
To establish this let ${\cal G}_r'$
be the subset of ${\cal G}_r$ with the following properties.
\begin{enumerate}\label{nontree}
\item [(P1)] $G$ is connected, and the second eigenvalue of the adjacency
matrix of $G$ is at most
$2\sqrt{r-1}+\ve$, where $\ve>0$ is an
arbitrarily small positive constant.
\item [(P2)] Let $s = O(\log n)$, and let $a= \rdup{2s(\log re)/\log n}$.
No set  of vertices $S$ of size $s$
induces more than $s+a$  edges. In particular, for
 $s \le (\log n)/(2\log re)$  no set  of vertices $S$ of size $s$
induces more than $s$  edges.
\end{enumerate}

\begin{lemma}\label{typG}
Let ${\cal G}_r' \seq {\cal G}_r$ be the $r$-regular graphs satisfying (P1), (P2). Then
$|{\cal G}_r'| \sim |{\cal G}_r|$.
\end{lemma}
\proofstart
Friedman \cite{F}, shows the deep result that (P1) holds \whp\ for random
regular graphs. That (P2) holds \whp\ is straightforward to establish.
\proofend

\ignore{
{\em Proof of Corollary \ref{rreg},
 Mixing time etc.}
Assuming (P1), and  $r\ge 4$, we can
take $\ve =1/10$. This implies that
$\F_G \ge 1/20$, and we can choose some large constant $B$ such that
\begin{equation}\label{Tlogn}
T = B\log n,
\end{equation}
satisfies \eqref{4}. As $\pi_v=1/n$ we have $T\pi_v=o(1)$.
}

{\em Proof of Corollary \ref{rreg}.}
Let $\ell= \e \log n$ for some $\e>0$. Property (P2) implies the graph is $\ell$-good as follows.
For any vertex $v$ of the graph $G$, let $U^*$ be the smallest non-trivial connected, even degree,  vertex induced
subgraph rooted at $v$. As $r\ge 4$, this subgraph contains at least two cycles. Let $|U^*|=k$, then $U^*$
induces at least $k+1$ edges.
By property (P2), no subgraph on $s=\e \log n$ vertices with $\e=1/(2\log r e)$
induces more than $s$ edges, and  we conclude that $|U^*|>s$.

\subsection{Removing the even degree constraint?}\label{odd-deg}

The only place in the proofs where the even degree condition matters is the proof of
Observation \ref{obs1}, that the walk on unvisited edges  terminates at its start vertex.
How important is the even degree constraint?

We consider  the experimental  evidence for  the performance of the
$E$-process on both even degree, and odd degree graphs. In our experiments   unvisited
 edges are chosen uniformly at random.
We generated graphs of size up to half a million vertices,
using the random regular graph generator from the \emph{NetworkX} package ({\tt http://networkx.lanl.gov/}) for
the programming language Python.  This package implements the Steger/Wormald approach, see \cite{SW}.
We  used Python's built-in random number generator which is based upon the Mersenne Twister.
Each data point is the average of five actual experiments.

\ignore{
\begin{figure*}
\centering
\epsfig{file=flies.eps}
\caption{A sample black and white graphic (.eps format)
that needs to span two columns of text.}
\end{figure*}
}

In Figure \ref{fig:covertime} we plot the
normalised cover time of the $E$-process, in the case where the
choice of unvisited edges is random. The
{\em normalised} cover time is the actual cover time divided by $n$, as a
function of $n$. Thus,
  linear functions of $n$ appear flat etc.
The labeling on the graphs is as follows:
The first letter  indicates an $E$-process, and this is followed by the degree $d=r$ of the graph.
In the case where the plot appears to be non-linear, a
curve of the form $c \log n$, is drawn behind the normalised experimental data, and labeled $[c n \ln (n)]$.
The constant
 $c$ used to draw the curve was determined by inspection.

It would appear  the plots for
 even degrees $4$ and $6$ are constant, i.e. the cover time is $O(n)$.
On the basis of experimental evidence,
the normalised cover time of $3$-regular graphs is $\om(n)$; see
Figure \ref{fig:covertime}.
This $\om(n)$ growth appears to be $0.93 n \log n$.
For  degrees $5$ and $7$ the plot also appears to grow logarithmically.
We note, however, that it is notoriously difficult to quantify such growth on the basis of finite $n$, and we make no claims other
than to present our experiments.

\ignore{
We give an intuitive argument
to suggest why the cover time is $\Om(n \log n)$ when $r$ is odd.
We use the notation blue walk, to mean the walk on unvisited edges, and red walk to mean
the random walk on visited edges.
When $r=3$ there is a set of isolated
vertices $I$  of expected size $|I| \sim n/8$, left behind by the random blue walk.
This can be seen as follows. Fix a vertex $v$, and
 assume that $v$ is tree-like to
some fixed depth. All but $o(n)$ vertices satisfy this condition \whp.
The
probability that a random blue walk turns away from
$v$ each time it visits $N(v)$ is $(1/2)^3$.
If this occurs then $v$ is at the center of an isolated  blue star $\{v,w,x,y\}$.
Let $I$ be the set of such stars.
By a coupon collecting
argument, it should take $\Om(n \log n)$ steps for the red walk to visit all of $I$.
}

\end{document}